# A single-frequency single-resonator laser on erbium-doped lithium niobate on insulator

TIEYING LI, KAN WU*, MINGLU CAI, ZEYU XIAO, HONGYI ZHANG, CHAO LI, JUNMIN XIANG, YI HUANG AND JIANPING CHEN

*State Key Laboratory of Advanced Optical Communication Systems and Networks, Department of Electronic Engineering, Shanghai Jiao Tong University, Shanghai 200240, China*
*\*Corresponding author: [kanwu@sjtu.edu.cn](kanwu@sjtu.edu.cn)*

**Erbium-doped lithium niobate on insulator (Er:LNOI) is a promising platform for photonic integrated circuits as it adds gain to the LNOI system and enables on-chip lasers and amplifiers. A challenge for Er:LNOI laser is to increase its output power while maintaining single-frequency and single (-transverse)-mode operation. In this work, we demonstrate that single-frequency and single-mode operation can be achieved even in a single multi-mode Er:LNOI microring by introducing mode-dependent loss and gain competition. In a single microring with a free spectral range of 192 GHz, we have achieved single-mode lasing with an output power of 2.1 µW, a side-mode suppression of 35.5 dB, and a linewidth of 1.27 MHz.**

Lithium niobate on insulator (LNOI) or thin-film lithium niobate has attracted great attention as it introduces strong mode confinement to the lithium niobate (LN) and can significantly enhance the electro-optic and nonlinear effects. Together with the low absorption loss and wide transmission window from 400 nm to 5 µm, LNOI has become a popular platform for photonic integrated circuits [1-12]. Various excellent works have been reported including supercontinuum and second-harmonic generation [2-5], electro-optical modulation [6-8], microcombs [9-12], etc.

To achieve a fully integrated photonic system on LNOI, the on-chip gain is a key requirement as it provides light source, compensates loss, and enables the large-scale integration of components with different functions. Although it has been a mature technology of doping rare-earth ions ($Yb^{3+}$ and $Er^{3+}$) to LN crystals for many years [13], direct doping rare-earth ions to LNOI can only achieve a relatively low concentration with a shallow depth [14, 15]. Recently erbium-doped LNOI has been successfully fabricated by binding an erbium-doped LN wafer to a holder wafer with so-called "ion-cutting" technology and then polishing the erbium-doped LN layer to a proper thickness [16]. On this Er:LNOI platform, a few technologies have been proposed to fabricate desired devices including focused ion beam (FIB) [17], chemical mechanical polishing (CMP) [18-20], dry etching [21-24], etc. Various lasers and amplifiers have been demonstrated. For amplifiers, a small-signal gain >5 dB/cm and a total gain (signal enhancement) >18 dB has been reported [23-26].

Er:LNOI lasers are also demonstrated based on microring and whispering gallery mode (WGM) microresonators [17-22]. Unfortunately, most of these lasers are operated in either multi-transverse mode or multi-longitudinal mode. Vernier effect with two coupled resonators shows the possibility of single-frequency and single (-transverse)-mode operation [20, 22]. However, matching the resonances between two resonators requires precise fabrication control and may introduce extra loss if a resonance mismatch occurs. As a result, the output power is still low, typically tens or hundreds of nano-Watts [17-22]. Electrical tuning of one resonator can be a solution but it causes extra power consumption. Moreover, the small mode area of the single-mode Er:LNOI waveguide limits the maximum power that the waveguide can support. If one increases the mode area, multi-transverse mode operation occurs. The single-frequency operation also requires a small resonator for a large free spectral range (FSR), which further limits the maximum output power that a laser can achieve. Therefore, it is highly desired that a new design of Er:LNOI laser can be proposed which can simultaneously achieve a single-frequency single-mode operation and relatively high output power.

In this work, we have demonstrated a single-frequency single-mode Er:LNOI laser with a single microring resonator. The waveguide in the resonator is designed 2 µm wide to allow higher optical power. Although the waveguide supports multiple transverse modes, the high-order modes have more overlap with the rough sidewalls of the waveguide and thus experience higher loss in the resonator. With the participation of gain competition, only fundamental $TE_{00}$ mode can lase. The microring resonator has an FSR of 192 GHz (~1.5 nm near 1531 nm) and only single-frequency operation is supported near the gain peak of 1531 nm. Experimentally, single-frequency single-mode lasing with an output power of 2.1 µW, a side-mode suppression of 35.5 dB, and a linewidth of 1.27 MHz has been achieved. Without the design of two coupled resonators, this work with a single resonator has a much lower requirement on fabrication accuracy. This work can be a promising solution for a single-frequency single-mode laser on LNOI and paves the way to a fully integrated photonic system on LNOI for applications including optical communications and optical computing.

The laser was fabricated on an erbium-doped LNOI wafer. Briefly, with ion-cutting technology, Er:LNOI wafer was prepared by bonding a z-cut erbium-doped LN wafer onto a holder wafer with 2-μm $SiO_2$ and 400-μm silicon. After polishing, a thin-film Er:LNOI with 600 nm thickness and $SiO_2$ cladding was obtained. The doping concentration of erbium ions is 1 mol %. The device was then fabricated with the dry etching method. A hard mask of Cr was first patterned with photoresist hydrogen silsesquioxane (HSQ). The pattern was then transferred to the LN layer by using inductive coupled plasma reactive ion etching (ICP-RIE). The etching depth of the waveguide is 410 nm and the angle of the sidewall is 75°. The scanning electron microscopic (SEM) image of the fabricated device is shown in Fig. 1(a). The microring resonator has a radius of 105 μm, corresponding to an FSR of 192 GHz.

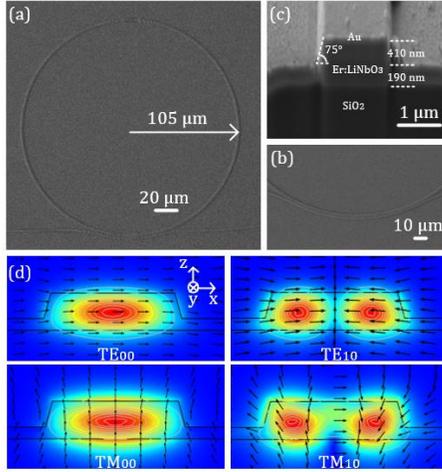

Fig. 1. (a). SEM image of microring with an FSR of 192 GHz. (b) Zoomed view of pulley coupling region of the microring. (c) The end facet after FIB polishing. (d) Four modes in the microring waveguide (The arrows indicate the direction of the electric field).

The microring is coupled to a bus waveguide with a pulley coupling region, as shown in Fig. 1(b). The waveguide of the microring is 2 μm wide and the bus waveguide is 1.4 μm wide. Fig. 1(c) shows an SEM image of the facet of the waveguide edge after FIB etching to improve the surface smoothness. Figure 1(d) shows four modes supported by the 2-μm wide waveguide. They are $TE_{00}$, $TE_{10}$, $TM_{00}$, and $TM_{10}$. As $TE_{10}$, $TM_{00}$, and $TM_{10}$ have large mode areas than $TE_{00}$, they overlap more with the rough sidewall and thus are expected to experience higher scattering loss. With the participation of gain competition in the Er:LNOI microring, only $TE_{00}$ mode is expected to lase.

The transmission spectrum of the microring resonator was characterized by a frequency-sweep laser (Agilent 8164A). To suppress the possible photorefractive and thermal effects in the Er:LNOI microring, the output power of the frequency-sweep laser was set to -5 dBm and a temperature controller was used to stabilize the chip temperature [18]. The measured transmission spectrum near 1530 nm is shown in Fig. 2(a). It can be seen that the baseline near 1 is not flat due to the weak amplified spontaneous emission (ASE) excited by the frequency-sweep laser. Meanwhile, the transmission dips related to multiple transverse modes can be observed clearly. By comparing the simulated and experimentally measured group indices ($n_g$) of different modes, the relation between modes and dips has been identified, as denoted in Fig. 2(a).

For the fundamental $TE_{00}$ mode, the linewidth of the transmission dip is 7.2 pm near 1531.49 nm, as shown in Fig. 2(b), corresponding to a loaded Q factor of $2.13 \times 10^5$. The simulated $n_g$, experimental FSR, $n_g$, and mode loss in the microring are summarized in Table 1. $TE_{00}$ mode has the lowest loss in the cavity, which is consistent with our analysis on the scattering loss from the waveguide sidewall.

The transmission spectrum near the pump wavelength of 1484 nm was also measured, as shown in Fig. 2(c). The linewidth near 1484.45 nm is 16.5 pm, corresponding to a loaded Q factor of $0.89 \times 10^5$, as illustrated in Fig. 2(d). The pump laser is a homemade Raman fiber laser. Its optical spectrum is also plotted in Fig. 2(c). It can be noted that only a small part of the pump power was coupled to the microring which limits the pumping efficiency of the Er:LNOI laser. In the future, a pump laser with narrower bandwidth can be adopted to improve the pumping efficiency.

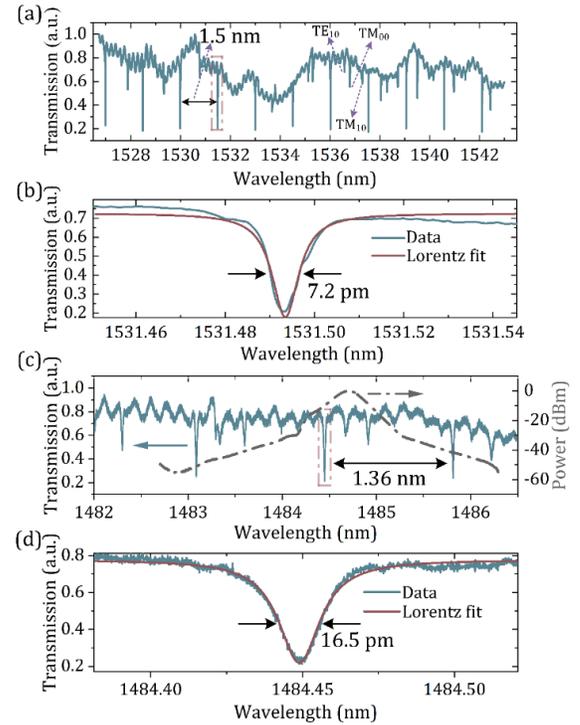

Fig. 2. (a) Transmission spectrum of microring from 1527 nm to 1543nm (b) Zoomed view of the resonance dip at lasing wavelength of 1531.49 nm. (c) Transmission spectrum of microring from 1482 nm to 1486.5 nm (green) and output spectrum of the pump laser (grey). (d) Zoomed view of the resonance dip near pump wavelength.

Table 1 Simulated $n_g$, experimental FSR, $n_g$, linewidth, and loss of different modes in the microring

| Modes | $TE_{00}$ | $TE_{10}$ | $TM_{00}$ | $TM_{10}$ |
|---|---|---|---|---|
| Simulated $n_g$ | 2.361 | 2.445 | 2.386 | 2.517 |
| FSR (GHz) | 192.225 | 185.229 | 190.068 | 179.852 |
| Experimental $n_g$ | 2.365 | 2.454 | 2.392 | 2.527 |
| Linewidth (pm) | 7.2 | 53.4 | 17 | 61.2 |
| Loss (dB/cm) | 1 | 11.4 | 3.2 | 13.2 |

The experimental setup of the Er:LNOI laser is shown in Fig. 3. The pump light from the 1484-nm pump laser first propagated through a fiber polarization controller (PC) and was then guided to the bus waveguide on the chip via a lensed fiber. The single-ended coupling loss is 6 dB. The output of the Er:LNOI laser was collected by another lensed fiber. The residual 1484-nm pump was extracted by a wavelength division multiplexer (WDM) and the signal near 1531 nm was measured by an optical spectrum analyzer (OSA). The inset in Fig. 3 shows the fluorescence micrograph when the 1484-nm pump is injected into the chip.

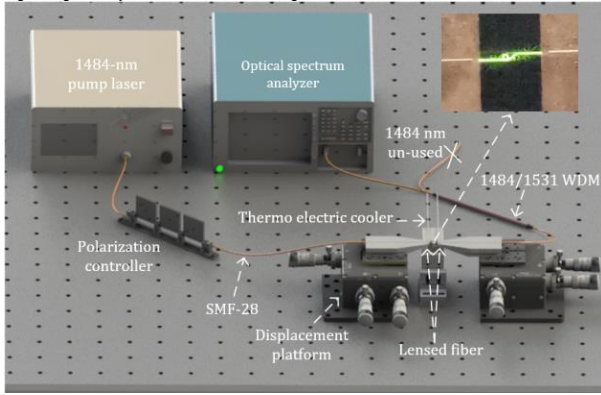

Fig. 3. Experimental setup for the Er:LNOI microring laser. WDM: Wavelength Division Multiplexing, Inset: Micrograph of the chip pumped at 1484 nm.

The measured optical spectrum is shown in Fig. 4(a). It can be observed clearly that only one wavelength near 1531 nm is lasing. The side-mode suppression ratio is 35.5 dB. A zoomed view near 1531 nm is shown in Fig. 4(b). Small peaks of different modes are denoted. Three transverse modes ($TE_{10}$, $TM_{00}$, $TM_{10}$) have been successfully suppressed. To explain the suppression of longitudinal modes, we measured the ASE spectrum from another straight Er:LNOI waveguide, as shown in Fig. 4(c). It can be seen that there is a sharp gain peak near 1531 nm which means the gain filtering effect can help to suppress the longitudinal modes far away from 1531 nm. On the other hand, these longitudinal modes ($TE_{00}$) still experienced less loss compared to those transverse modes ($TE_{10}$, $TM_{00}$, $TM_{10}$). Therefore, the powers of these longitudinal modes determined the SMSR. As shown in Fig. 4(a), the highest peaks of these suppressed longitudinal modes appear near the lasing wavelength of 1531.3 nm and the second gain peak near 1545 nm. The output spectra under different on-chip pump power are shown in Fig. 4(b) and the output power values are summarized in Fig. 4(d). The threshold pump power is 14.5 mW. It should be emphasized that this on-chip pump power is the pump power in the input fiber subtracting the 6-dB input coupling loss. The actual pump power coupled to the microring is much lower because of the resonance condition of the microring and the absorption from the bus waveguide before the microring. The actual pump power coupled to the microring is estimated to be 20.3 dB lower than the on-chip pump power, which is obtained by comparing the pump spectra before and after (inset in Fig. 4(a)) the microring (19.7 dB) and estimating the absorption from the bus waveguide (0.6 dB). At an on-chip pump power of 32 mW, the laser output power can reach up to 2.1 μW, which is already nearly one order higher than that reported currently [17-22]. The slope efficiency of the chip laser is $1.20\times10^{-4}$.

It is noted that the lasing wavelength in our work blue shifted to a shorter wavelength with the increase of pump power, as shown in Fig. 4(e). This is different from the reported works in which the lasing wavelength redshifted due to the thermal effect in the resonator when the pump power increased [17], The output wavelength is nearly linear with the pump power with a slope of -2.87 pm/mW, as shown in Fig. 4(f). The corresponding total index change with 12.5-mW increase of on-chip pump power is $-4.68\times10^{-5}$. The actual increase of pump power in the microring is 12.5 mW × 0.87 × 14.35 = 156.0 mW where 0.87 is the -0.6 dB loss from the bus waveguide before the microring and 14.35 is the cavity enhancement obtained by fitting the resonance dip at 1484 nm in Fig. 2(d). The power-dependent index change in the microring is therefore calculated to be $-3.0\times10^{-4}$ W$^{-1}$.

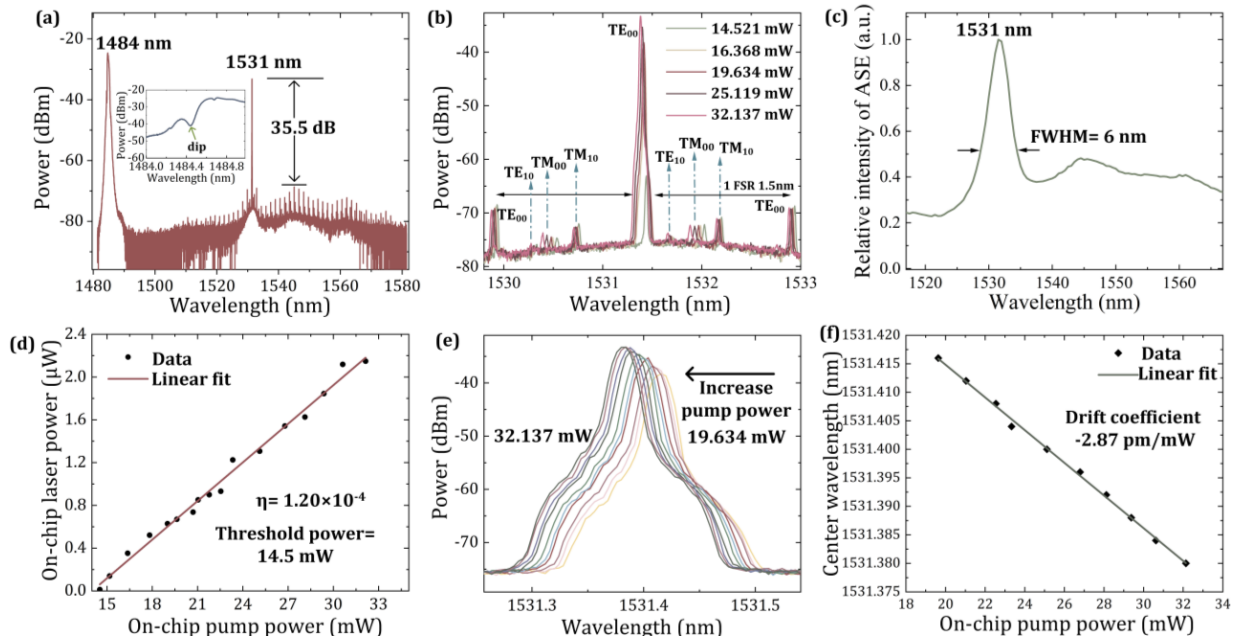

Fig. 4 (a) Output spectrum of the microring laser. Inset: zoomed view of the pump spectra after passing through the microring. (b) Spectral evolution of the microring laser with increased pump power. (c) ASE of an Er:LNOI straight waveguide. (d) On-chip output power versus on-chip pump power. (e) Spectra of central wavelength shift with increased pump power. (f) Central wavelength versus on-chip pump power.

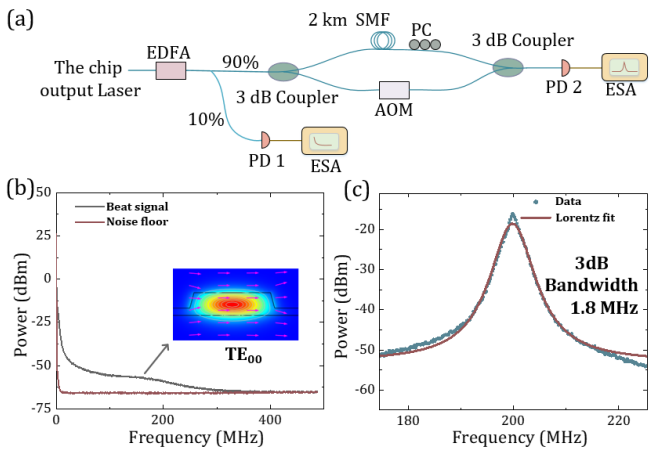

Fig. 5 (a) Experimental setup for confirmation of single-frequency single-mode operation and linewidth measurement. The beating signals are shown for (b) PD 1 and (c) PD 2. Inset in (b): $TE_{00}$ mode profile. EDFA: erbium-doped fiber amplifier, PD: photodetector; ESA: electrical spectrum analyzer. AOM: acousto-optic modulator. PC: polarization controller SMF: single-mode fiber

This negative relation between the index change and pump power is attributed to the photorefractive effect [27] where the refractive index decreases with the increase of light power. The influence of two other effects, i.e. thermal effect and active medium induced index change, have negligible influence on our device. For the thermal effect, the electric field of the laser ($TE_{00}$) in the waveguide oscillates in $n_o$ direction due to the z-cut LN. The thermo-optic coefficient in this direction $\Delta n_o/\Delta T$ is $-6\times10^{-8}$ K$^{-1}$ at a room temperature of 298 K which contributes a tiny index change of $\sim10^{-6}$ with a temperature change of 10~20 K [28]. For the active medium induced index change, the $\Delta n/\Delta P$ for $Yb^{3+}$ (or $Er^{3+}/Yb^{+3}$) doped fiber is $3.3\times10^{-6}$ W$^{-1}$ (or $8.4\times10^{-7}$ W$^{-1}$) [29] which is two orders smaller than the measured value of $-3.0\times10^{-4}$ W$^{-1}$ in our device. We estimate this effect would not significantly change in Er:LNOI. Thus, the photorefractive effect is believed to dominate the blue shift of cavity resonance in our device.

To confirm the single-frequency single-mode operation and to measure the linewidth of our laser, the experiment in Fig. 5(a) was performed. The laser output was first amplified by an erbium-doped fiber amplifier (EDFA). Then 10% of its power was directly detected by a photodetector (PD1) and characterized by an electrical spectrum analyzer (ESA). The results are shown in Fig. 5(b). It can be seen that no beating signal is observed within a 500 MHz span. 90% of the power was guided to a self-homodyne detection setup for the linewidth measurement [30]. The acousto-optic modulator (AOM) shifted the light frequency by 200 MHz. The linewidth measurement result is shown in Fig. 5(c). A Lorentz fit curve is also plotted. The calculated linewidth is 1.27 MHz.

In conclusion, we have demonstrated an Er:LNOI microring laser with a single-frequency single-mode operation. With the mode-dependent loss and gain competition, only a single $TE_{00}$ mode oscillates in a multi-mode microring cavity. The laser has a maximum output power of 2.1 μW, a side-mode suppression ratio of 35.5 dB, and a linewidth of 1.27 MHz. Our work paves the way for the development of on-chip Er:LNOI laser and the realization of large-scale photonic integrated circuits on LNOI in the future.

**Funding.** National Natural Science Foundation of China (NSFC) (61922056, 61875122).

**Disclosures.** The authors declare no conflicts of interest.